\documentstyle[12pt]{article}
\def\cc{\hbox{C\kern-0.55em\raise0.4ex\hbox{$\scriptstyle\bf |$}}}
\def\lo{\raise0.3ex\hbox{$<$\kern-0.75em\raise-1.1ex\hbox{$=$}}}
\newcommand{\be}{\begin{eqnarray}}
\newcommand{\ee}{\end{eqnarray}}
\newcommand{\no}{\nonumber}

\begin{document}
\sl
\begin{titlepage}
\rightline{FTUV/93-13}
\hspace*{2.0cm}
\begin{center}
{\bf MOTION OF WAVEFUNCTION ZEROS IN SPIN-BOSON SYSTEMS}\\
\\
\ \\
{\bf Demosthenes Ellinas$^{1)}$ and Vassilios Kovanis$^{2)}$}
\vskip 0.5 cm
$^{1)}$
Departamento de Fisica  Teorica, Faculdad de Fisica \\ Universidad de Valencia,
 E-46100 Burjasot,
Valencia, Spain \\
{\it ellinas@evalvx.ific.uv.es}
\vskip 0.5 cm
$^{2)}$Nonlinear Optics Center , Phillips Laboratory,\\ 3350 Aberdeen SE,
Kirtland AFB, NM 87117-5776, USA\\
{\it kovanis@xaos.plk.af.mil}
\end{center}
\vskip 2.0 cm
\begin{abstract}
In the analytic-Bargmann representation associated with the harmonic
oscillator and spin coherent states, the wavefunction as entire complex
functions can be factorized in terms of their zeros in a unique way.
The Schr\"odinger equation of motion for the wavefunction is turned to
a system of equations for its zeros. The motion of these zeros as a non-linear
flow of points is studied and interpreted for linear and non-linear bosonic
and spin Hamiltonians. Attention is given to the study of the zeros of
the Jaynes-Cummings model and to its finite analoque. Numerical solutions
are derived and discussed.
\end{abstract}
\end{titlepage}
{\bf I. INTRODUCTION}
\vskip 0.5 cm
Coherent states constitute a basic tool in quantum optics as well as in
many other branches of quantum physics [1]. Starting with the early attempt
of Schr\"o\-dinger [2] to construct a wavepacket as-classical-as-possible
and later on with Glauber [3] in the statistical studies of light radiation,
the
coherent states of the harmonic oscillator became the prototype for the
subsequent generalizations to the spin coherent states [4,5] and to
other more general algebraic sets [6]. A major step forward in the theory
of coherent states was taken by Bargmann [7] who associated to them the entiry
functions and provided the analytic representation of the creation and
annihilation
operators of the harmonic oscillator. Similarly to the mapping of the vectors
of
Hilbert space to holomorphic functions (polynomials of infinite degree in
$\alpha\in\cc\ )$, as is the case for the bosonic coherent state, in the case
of the
$j$-spin coherent states the $2j+1$ dimensional Hilbert space vectors are
mapped to entire polynomials of at most $2j$ degree in the $z\in\cc\ $
variable.

The
starting point of the present work is exactly this holomorphic and
polynomial character of the wavefunction in the Bargmann representation.
According to the theory of polynomials (the so called fundamental theorem
of algebra), a polynomial can be factorized in terms of its roots (zeros)
in a unique and faithful way. That is to mean that the mere knowledge of the
zer
o
locations on the complex plane enables the reconstruction of the wavefunction,
in a unique way. However care
must be exercised to the case where some degeneracy occurs between the
roots, as we will see later on in the text (Chapt. II). The root-factorization
scheme in the case of the ocillator
coherent states where the number of zeros of the entire wavefunction might
be infinite is also known as the
Weierstrass-Hadamard decomposition theorem [8] and will be adopted for our
case of the harmonic oscillator (Chapt. III).
Addressing next the problem of dynamics of the wavefunction we will
attribute its evolution to the motion of its roots and so we will transcript
the Schr\"odinger equation to a system of equations giving the motion of zero
locations on the complex plane . In the case of Hamiltonians linear in the
generators of the oscillator or the $SU(2)$ algebra the equations are in
general
of the form of a decoupled system of Riccati equations which are easy to
linearize
and treat analytically (Chapt. IIa, IIIa). However, for non-linear Hamiltonians
with the simple type of non-linearities occurring in e.g. the Raman-Nath
Hamiltonians, the equations of zeros become a complex non-linear system
which apparently can be solved only by numerical methods, (Chapt. IIb, IIIb).
Extending the previous considerations to more complicated systems we take up
the case of spin-bosons Hamiltonians, exemplified by the
Jaynes-Cummings model (JCM), (Chapt. V) and by what we call finite-JCM
or $SU(2)$-JCM, a model in where the boson operators of the original model
have been
replaced by the respective $SU(2)$ generators, (Chapt. IV). The cases
of rotating-wave-approximation (RWA) and non-RWA are treated separately
and the equations of motion of the zeros are derived. Simplification of
the ensuing involved equations is achieved by reduction to cases of only
one and two zeros. Such cases are treated numerically and the flow of zeros
on the plane is obtained and discussed.

Finally we should mention that root-factorization schemes similar to
those of this paper occur in the studies of motion of charged particles in
magnetic fields [9] e.g. Hall effect [10,11] and also recently the spin
coherent states factorization has been used in the study of chaotic dynamics
of kicked tops [12,13,14].
\vskip 1.0 cm
{\bf II. THE SPIN CASE}
\vskip 0.5 cm
Let us start the treatment of the spin case by first recalling the
definition of the coherent state (CS),
\be
|z)=e^{\bar{z}J_+}|-j>=\sum\limits^{2j}_{m=0}(^{2j}_{m})
^{\frac{1}{2}}\bar{z}^{m}|-j+m>\ ,
\ee
which is normalized $|z>=\frac{1}{\sqrt{(z|z)}}|z)$, through the overlap
$(z|z)=(1+|z|^2)^{2j}$. Since we want to use the analytic properties the
CS we will employ in the sequel the unormalized state of eq. (1)
which is an analytic polynomial of $2j$ degree. The normalized CS is
non-analytic in $z$ due to the normalization factor and so not suitable
for our purposes. Now, any vector of the $2j+1$-dimensional repsentation
space of the spin written by $|\psi>=\sum\limits^{2j}_{n=0}c_n|n>$, has
components along the CS basis
\be
\psi (z)=(\bar{z}|\psi>=\sum\limits^{2j}_{m=0}c_n
(^{2j}_{m})^{\frac{1}{2}}z^{m}\ ,
\ee
where the overbar denotes complex conjugation. The analytic function
$\psi(z)$ is a complex polynomial of at most $2j$ degree. In this analytic
representation of the Hilbert space the monomials $\gamma_{jm}(z)
\equiv(^{2j}_{m})^{\frac{1}{2}}
z^{m}$, constitute a orhonormal basis with inner product $(z=re^{i\theta})$,\\
$<\bar{\gamma}_{jn}^{(z)},\gamma_{jm}^{(z)}>=$
\be
\frac{(2j+1)}{\pi}\int d^2z\frac{\bar{\gamma}_{jn}(z)\gamma_{jm}(z)}
{(1+z\bar{z})^{2j+2}}=\delta_{nm}2(2j+1)\int^\infty_0\ dr\frac{r^{2m+1}}
{(1+r^2)^{2j+2}}(^{2j}_{m})=\no\\
=\frac{(2j+1)!}{(2j-m)!m!}B(m+1,2j+1-m)=\delta_{mn}
\ee
and $B(k,\ell)$ stands for the Beta function,
\be
B(k,\ell)=\int\limits^\infty_0\frac{dx\ x^{k-1}}{(1+x)^{k+\ell}}=
\frac{\Gamma(\ell)\Gamma(k)}{\Gamma(\ell+k)}\ .
\ee
Similarly to the vectors the analytic realization of any spin operator
${\cal O}$ also is found by the formula
\be
(\bar{z}|{\cal O}|\psi>&=&<-j|(e^{zJ_-}{\cal O}e^{-zJ_-})e^{zJ_-}|\psi>=\no\\
&=&<-j|({\cal O}+z[J_-,{\cal O}]+\frac{z^2}{2}[J_-,[J_-,{\cal O}]]+\cdots
)e^{zJ_-}|\psi>\ .
\ee
Depending on the kind of commutation relation each operator $\cal O$ has
with $J_-$ the above series terminates after some terms or a summable
series of terms is obtained. In this manner
we found the following realizations $(\partial_z\equiv\partial/\partial z)$:
\be
J_-&=&\partial_z\ ,\hspace{5.0cm}\no\\
J_+&=&2jz-z^2\partial_z\ ,\hspace{5.0cm}\no\\
J_0&=&-j+z\partial_z\ ,\hspace{5.0cm}\\
{\rm and}\hspace{4.0cm}\no\\
J^2_0&=&j^2+(-2j+1)z\partial_z+z^2\partial^2_z\hspace{5.0cm}\no\ .
\ee
In the above realization the operators $J_+$ and $J_-$ are adjoint to
each other, since for any two analytic functions
$\psi_a(z)=\sum\limits^{2j}_{m=
0}
a_mz^m$ and $\psi_b(z)=\sum\limits^{2j}_{m=0}b_mz^m$, is valid that $<\psi_a,
J_-\psi_b>=<J_+\psi_a,\psi_b>$ by virtue of the following proof:
\be
<\psi_a,J_-\psi_b>=\frac{(2j+1)}{\pi}\int\frac{d^2z}{(1+z\bar{z})^{2j+2}}
\bar{\psi}_a\partial_z\psi_b=\hspace{5.0cm}\no\\
(2j+1)\sum\limits^{2j}_{m=0}(m+1)\bar{a}_m
b_{m+1}\int\limits^\infty_0\frac{dr^2r^{2n}}{(1+r^2)^{2j+2}}=
\sum\limits^{2j}_{m=0}\bar{a}_mb_{m+1}\frac{(m+1)!(2j-m)!}{(2j)!}\ ,
\ee
and
\be
<J_+\psi_a,\psi_b>=\frac{(2j+1)}{\pi}\int\frac{d^2z}{(1+z\bar{z})^{2j+2}}
\psi_b\overline{(2jz-z^2\partial_z)\psi_a}=\hspace{3.5cm}\no\\
(2j+1)\sum\limits^{2j}_{m=0}
(2j-m)b_{m+1}\bar{a}_m\int\limits^\infty_0\frac{dr^2r^{2(n+1)}}{(1+r^2)
^{2j+2}}
=\sum\limits^{2j}_{m=0}\bar{a}_m b_{m+1}\frac{(2j-m)!(m+1)!}{(2j)!}\ .
\ee
We turn now to the factorization theorem, which as was outlined about states
that the analytic wavefunction $\psi(z)$ as a polynomial of degree $2j$, is
uniquely characterized by its zeros,
\be
\psi(z)=\prod\limits^{2j}_{m=1}(z-z_m)\ .
\ee
As the variable $z=e^{-i\theta}\cos\frac{\theta}{2}$, gives the complex
stereographic coordinate of a point $(\phi,\theta)$ on the sphere, there
are two ways to depict the zeros $z_m$'s, either as points on the tangent
plane of the sphere on the north pole or as points on the surface of the
sphere. Following the latter representation which was originally introduced
by Majorana [15], discussed further by Bacry [16] and has been reviewed
in, e.g. Ref.17, we will
speak about a constellation of stars (zeros) which uniquely corresponds
to each wavefunction. To each state $|m>$, correponds a degenerate
constellation of $m$ stars on the same point of the sky. To the ground
state corresponds the dark sky while to a coherent state $|z_0)$,
corresponds a $2j$-degenerate star at the point $-1/\bar{z}_0$.

We can
exploit the factorization theorem to study also the dynamics of the
wavefunction which will be attributed entirely to the dynamics of its
zeros. If so, then the analytic realization of the Schr\"odinger equation
(assuming $\hbar=1$), $i(z|\dot{\psi}>=(z|H(J_{\pm,0})|\psi>$ which is written
as $i\dot{\psi}(z)={\cal H}(z,\partial_z)\psi(z)$ is equivalent to a
system of $2j$ first order differential equations for the zeros. Indeed if
$\psi(z_k;t)\approx 0$ at some time $t$, then by demanding that at a later
time $\psi(z_k+\delta_{z_k}\ ;\ t+\delta_t)\approx 0$ we obtain by Taylor
expansion, $\partial_z\psi(z)\biggl|_{z_k,t}\delta z_k+\partial_{t'}\psi(z)
\biggl|_{z_{k,t}}\delta t\approx 0$, which implies that [13]
\be
\dot{z}_k\equiv\frac{\delta z_k}{\delta t}=-\frac{\partial_t\psi(z)}
{\partial_z\psi(z)}\left|\begin{array}{l}\ \\ z_k\end{array}\right.
=-i\frac{{\cal H}(z,\partial_z)
\psi(z)}{\partial_z\psi(z)}\left|\begin{array}{l}\ \\ z_k\end{array}\right.
\ee
for $k=1,2,...,2j$. This is the equation which provides together with the
factorization issued in eq. (9) the motion of the zeros on the complex plane
which may be represented as a flow of points. The kind of this motion and
the mutual interactions of the zeros, evidently depend on the Hamiltonian
and especially on the highest power of the derivative $\partial_z$.
Hamiltonians linear in the generators are by virtue of eq. (6) first order
in $\partial_z$ and therefore the motion of each zero is decoupled from
the others. While non-linear Hamiltonians induce interactions between
the zeros which now execute more complicated motions  on the complex plane.
In the next section we will elucidate this by two characteristic examples,
but before that we shall briefly elaborate on the case with degenerate zeros.
The factorization in eq. (9) assumes no degeneracy between the zeros. If
instead we regard the case where $\ell$ out of $2j$ roots are non-degenerate
$z_1,z_2,\cdots,z_\ell$ and the rest $2j-\ell$ are degenerate i.e. $z'_1=z'_2=
\cdots =z'_{2j-\ell}=z'$, then $\psi(z)=(z-z')^{2j-\ell}\prod\limits^\ell_{n=1}
(z-z_n)$. Then we readily find that the first order derivative, in the
Taylor expansion, of the wavefunction, at the degenerate roots is zero
and if the speed of motion of the zeros is going to assumed finite then
higher order derivatives must be considered and this leads to second order
time derivatives for the motion of zeros. Such generalized situations are
worth exploring further since they will probe the dynamics of wavefunction
with a number of degeneracies in their roots, here instead we shall confine
ourselves to the case of multiplicity free zeros.
\pagebreak

\begin{center}
{\bf A. Linear spin}
\end{center}
\vskip 0.3 cm

If we consider a linear Hamiltonian in the generators of the spin algebra,
i.e. $H=a_3J_3+a_+J_++a_-J_-$ with analytic realization
\be
{\cal H}=-a_3j+a_+2jz+(a_3z-a_+z^2+a_-)\partial_z\ ,
\ee
which physically could account for the motion of an arbitrary spin
in a magnetic field, we find from the general equation of motion (10) and
the factorization (9) that each zero of the wavefunction satisfies
the following Riccati equation,
\be
\dot{z}_k=i(a_3z_k-a_+z^2_k+a_-)
\ee
with $k=1,2,...,2j$. This is linearizable equation and there are standard
textbook solutions in the case of
time independent coefficients and for some classes of time
dependent coefficients as well. Interestingly the same equation occurs
in other methods of solving the problem and especially in the path integral
approach the evolution of the propagator amounts to solving such a Riccati
equation [18]. What is more interesting here is that the linearity of
Hamiltonian
entails a separate motion of each zero which implies further that the
structure of the root decompostition of the initial wavefunction is not
changed during the evolution. Especially the possible degeneracies of the
wavefunction at the start of the evolution are not lifted. This implies
that a initial coherent state with a $2j$-fold degeneracy evolves
into a new coherent state due to the preservation of the degeneracy of
its corresponding constellation of zeros.
\vskip 0.3 cm
\begin{center}
{\bf B. Non-linear spin}
\end{center}
\vskip 0.3 cm

To study the effect of non-linearity on the motion of zeros we use the spin
Raman-Nath Hamiltonian[19,20]
\be
H=aJ^2_0+\frac{b}{2}(J_++J_-)
\ee
which physically accounts for the transverse motion of a particle in a standing
wave created by two counterpropagated laser fields. The analytic realization
of the Hamiltonian is by virtue of eq. (6) a second order polynomial in
the complex derivative:
\be
{\cal H}=a[j^2+(-2j+1)z\partial_z+z^2\partial_z^2]+\frac{b}{2}(\partial_z+
2jz-z^2\partial_z)\ .
\ee
We then obtain the equations of motion,
\be
\dot{z}_k=i a\biggl[ z^2_k\sum\limits^{2j}_{k\neq i=1}
\frac{1}{(z_k-z_i)}-\biggl( j-\frac{1}{2}\biggr) z_k\biggl]-i\frac{b}{2}
(z^2_k-1)\ .
\ee
The non-linearity introduces a coupling between the motion
of the zeros which now are not evolving independently as in the linear case,
and moreover due to the sum term above, any degeneracy of zeros is
lifted in the course of the evolution.  In fact the sum term acts as a
repeller of zeros and is responsible for the non-conervation of an
initial coherent state. This repeller is due to the non-linear term
in the Hamiltonian and its lifting of degeneracy of a constellation
accounts  geometrically for the well know fact that under non-linear
Hamiltonians the spin coherent state is unstable, since it is not
evolving to a new coherent state. It is also worth noticing that due to the
non-linearity there appear in the r.h.s of eq. (15) the $j$ index, which
makes the dynamics of zeros to dependent on the representation of the spin
as well. This dependence was absent of course in the linear case.
\vskip 1.0 cm
{\bf III. THE HARMONIC OSCILLATOR CASE}
\vskip 0.5 cm
For the harmonic oscillator the coherent states and the Bargmann analytic
repsesentation is a fairly standard material[1,6,7] which for later use
will only briefly
summarized here before we come to the factorization theorem.

The coherent state $|\alpha)=e^{\bar{\alpha}a^+}|0>$ gives a holomorphic
function
with respect to the complex $\bar{\alpha}$. The (over)completeness of the
coherent state basis is used to map any arbitrary vector $|\psi>$
of the Hilbert space spanned by the number states, $|\psi>=\sum\limits^\infty
_{n=0}c_n|n>$, to a holomorphic function,
\be
\psi(\alpha)=(\bar{\alpha}|\psi>=\sum\limits^\infty_{n=0}c_n(\bar{\alpha}|n>=
\sum\limits^\infty_{n=0}c_n\alpha^n\ .
\ee
In this analytic representation the monomials $\gamma_n(\alpha)\equiv
\frac{\alpha^n}{\sqrt{n!}}$, constitute a orthonormal basis with Bargmann
measure of integration $e^{-|\alpha|^2}$ according to the following:
\be
<\bar{\gamma}_m(\alpha),\gamma_n(\alpha)>=\frac{1}{\pi}\int d^2\alpha
e^{-|\alpha|^2}\frac{\alpha^m}{\sqrt{m!}}\frac{\alpha^n}{\sqrt{n!}}=
\delta_{mn}\ .
\ee
As in the spin case we utilize eq. (5), together with the commutation
relations of the oscillator algebra to find the analytic realization of
any operator. We will need only the following ones:
\be
a&=&\partial_\alpha\ ,\hspace{6.0cm}\no\\
a^+&=&\alpha\ ,\hspace{6.0cm}\no\\
N&=&\alpha\partial_\alpha\ ,\hspace{6.0cm}\\
{\rm and}\hspace{5.0cm}\no\\
N^2&=&\alpha\partial_\alpha+\alpha^2\partial^2_\alpha\ .\hspace{6.0cm}\no
\ee
We should mention that with respect to the inner product of eq. (17)
the realizations (18) of the creation and annihilation operators
are adjoint to each other, i.e. $<\psi_a,a\psi_b>=<a^+\psi_a,\psi_b>$,
for any two arbitrary holomorphic functions $\psi_{a,b}$.

Coming now to the factorization suitable for the entire wavefunctions $\psi
(\alpha)$, pertinent to the present case, we can
regard these entire functions as generalizations of the polynomials
which however have an infine number of (isolated) zeros. According to a
known theorem in the theory of complex functions, the so called
Weierstrass-Hadamand theorem [8], the root-factorization of an entire
function is unique only up to a factor. This factor is an entire function
which can be specified from the knowledge of the growth properties
of the function in question.
The growth properties of a given holomorphic function $\psi(\alpha)$,
characterize the growth rate of the magnitude of the function $|\psi(\alpha)|$,
when $|\alpha|\equiv r\rightarrow\infty$. As a measure of this growth rate
we may take the exponential function $e^{\tau r^\rho}$. If the growth rate
of the function is limited so that when $r\rightarrow\infty$,
\be
|\psi(\alpha)|\lo e^{\tau'r^{\rho'}}\ ,
\ee
then if $\rho$ and $\tau$ are the minima between the $\rho'$'s and $\tau'$'s
for which the above condition holds, we will say that the function is of
order $\rho$ and of type $\tau$. Then according to the aforementioned
Weierstrass-Hadamand theorem a holomorphic function of growth $(\rho,\tau)$
can be decomposed in terms of a product over the infinite set of zeros
$\alpha_m$, such that $\psi(\alpha_m)=0$, as
\be
\psi(\alpha)=a^\nu e^{a_0+a_1\alpha +\cdots +\alpha^p}\prod^\infty_{m=1}
(1-\alpha/\alpha_m)e^{\frac{\alpha}{1}+\frac{\alpha^2}{2}+\cdots +
\frac{\alpha^p}{p}}
\ee
where $p\leq\rho$, and the $a$'s are arbitrary coefficients and a
$\nu$-fold zero is
assumed to be located at the origin. To employ this decomposition
in our case of the oscillator related entire functions we should specify
first the growth properties of the wavefunction. This is done by the
square-integrability condition for the wavefunctions i.e.:
\be
<\psi|\psi>=\parallel |\psi>\parallel^2=\frac{1}{\pi}\int d^2\alpha
e^{-|\alpha|^2}|\psi(\alpha)|^2<\infty\ ,
\ee
which suggests that the wavefunction must be of growth $0<\rho\leq 2$ and
$\tau=\frac{1}{2}$.

This implies that for the wavefunction of order $\rho=2$ assuming for
simplicity that there are no zeros at the origin the root-decomposition
becomes,
\be
\psi(\alpha)=e^\alpha\prod^\infty_{m=1}(1-\frac{\alpha}{\alpha_m})
e^{\alpha/\alpha_m}=e^\alpha\prod^\infty_{m=1}E(\frac{\alpha}{\alpha_m})
\ee
where the abbreviation $E(\frac{\alpha}{\alpha_m})$, known as Weierstrass
primitive factor has been used. In comparison with the spin case we have
here the appearance of an additional exponential factor in the factorization,
which guarantees the convergence of the infinite product.
\pagebreak

\begin{center}
{\bf A. Linear oscillator.}
\end{center}
\vskip 0.5 cm

To probe the dynamics we will first apply the equation of motion (10), to the
simplest case of a linear oscillator with a classical drive, with Hamiltonian,
$H=\omega N+\gamma a^++\bar{\gamma}a$. From the derivative realizations of
eq. (18), we easily derive the equation of zeros,
\be
\dot{\alpha}_m=-i\omega\alpha_m\ +\bar{\gamma}
\ee
for $m=1,2,...,\infty$. Each zero of the wavefunction is decompled from the
rest and executes a rotation on the complex plane driven by the coupling
constant. This independence in the motion of zeros implies, as in the linear
spin case, a preservation of an initial coherent state. This simple picture is
however greatly changed in the case of non-linear oscillator as we see next
for the Raman-Nath Hamiltonian.
\vskip 0.5 cm
\begin{center}
{\bf B. Non-linear oscillator}
\end{center}
\vskip 0.5 cm

A quadratic non-linearity in the number operator is the characteristic of the
bosonic Raman-Nath equation will as we mentioned above accounts for deflection
of atomic [19,20] and light beams when crossing standing waves [21]. Its
Hamiltonian generically reads,
\be
H=\lambda N^2+\gamma a+\bar{\gamma}a\ ,
\ee
then turning the Schr\"odinger equation of the wavefunction to equations for
its zero is a lengthy straightforward procedure which yields,
\be
\begin{array}{l}
\dot{\alpha}_m=-i\{\lambda\alpha^2_m(e^{\alpha_m}-1)-\lambda\alpha_m
+\bar{\gamma}+\\
\ \\
+\lambda\alpha_m\frac{\sum\limits_{\ell}\prod\limits_{\ell\neq k}
2E(\frac{\alpha_m}{\alpha_\ell})E(\frac{\alpha_m}{\alpha_k})
-e^{\frac{\alpha_m}{\alpha_\ell}-\frac{\alpha_m}
{\alpha_k}}E(\frac{\alpha_m}{\alpha_k})-e^{\frac{\alpha_m}{\alpha_k}-
\frac{\alpha_m}{\alpha_\ell}}E(\frac{\alpha_m}{\alpha_\ell})-(\frac{1}
{\alpha_\ell}+1)E(\frac{\alpha_m}{\alpha_\ell})-e^{-2\frac{\alpha_m}
{\alpha_\ell}}}{\sum\limits_\ell\prod\limits_{\ell\neq k}
[E(\frac{\alpha_m}{\alpha_k})-1]E(\frac{\alpha_m}{\alpha_\ell})}\}\end{array}
\ee
We observe from the above equation that the quadractic nonlinearity of the
Hamiltonian amounts to a highly non-linear coupling between the zeros which
takes place through the exponential terms and by the product factors. The
latter of course converge rapidly and only few of their elements contribute
significantly, still however the equations apparently are not possible to
tackle by any analytic method. One could only expect to study the flow of
zeros numerically in some low photon-number excitation limit where the
wavefunction can be truncated to a finite polynomial of low degree. Such
cases will be treated below where Hamiltonians with
coupled spin and bosonic degrees of freedom are considered.
\vskip 1.0 cm
{\bf IV THE FINITE JAYNES-CUMMINGS-MODEL CASE}
\vskip 0.5 cm
Let us extend now our study to models involving coupled spin and bosons
operators. Our
aim is to treat a single bosonic mode in coupling with a spin or a two-level
atom. We will use the simple case of a linear spin-boson interaction as in
the JCM to probe the effect of the spin onto the dynamics of the zeros of
the wavefunction corresponding to the mode. For this reason the atomic
operators will be treated as two-dimensional matrices written in terms of
the Pauli matrices, and for the boson mode we will employ the analytic
realization of its generators.

Motivated by the simplicity of the root-decomposition for the spin compared
to the similar, but a lot more complicated one pertinent to the bosons, as
became apparent from the previous sections, in this section we study a
finite analogue of the original JCM [22], in where the boson creation and
annihilation operators which generate the oscillator algebra, have been
substituted by the creation and annihilation generators of the $SU(2)$ rotation
algebra. Such a variant of the original JCM, called finite or $SU(2)$-JCM,
has been introduced and studied previously [23] in the context of quantum
phase operators. Using the method of group contruction it was shown in Ref.23,
that by appropriate scaling of the $SU(2)$ generators and the coupling constant
of the Hamiltonian the finite - JCM constitute a viable alternative to the
origin model, capable to reproduce in the contraction (high spin) limit
all the subtleties of the JCM dynamics, namely Rabi nutations and even
collapses
and revivals [24].

Since we will be interested in studing the system separately in the
rotating-wave-approximation (RWA) and in non-RWA we do not transform to a
rotating frame so on resonance the Hamiltonian is,
\be
H=\omega\sigma_3+\omega J_3+\lambda_1(\sigma^+J_-+\sigma^-J_+)+\lambda_2
(\sigma^+J_++\sigma^-J_-)
\ee
where $\omega$ is the common atom-field frequency and the Pauli matrices
are in coupling with the $SU(2)$ generators $J_\pm$ and finally we have RWA
when $(\lambda_1,
\lambda_2)=(\lambda,0)$ and non-RWA when $(\lambda_1,\lambda_2)=(\lambda,
\lambda)$. The equation of motion $i\vert\dot{\psi}>=H\vert\psi>$ has the form
\be
i\left(\begin{array}{l}
\vert\dot{\psi}^+>\\
\vert\psi^->\end{array}\right)=\left(\begin{array}{l}
H_{11}\ \ H_{12}\\
H_{21}\ \ H_{22}\end{array}\right)\left(\begin{array}{l}
\vert\psi^+>\\
\vert\psi^->\end{array}\right)=\\
\ \nonumber\\
\left(\begin{array}{ll}
\omega J_3+\omega & \lambda_1J_-+\lambda_2J_+\\
\lambda_1J_++\lambda_2J_- & \omega J_3-\omega\end{array}\right)\left(\begin
{array}{l}
\vert\psi^+>\\
\vert\psi^->\end{array}\right)\nonumber
\ee
Projecting along the $SU(2)$ coherent state $\vert z)$ we  obtain,
\be
i\left(\begin{array}{ll}
\dot{\psi}^+(z)\\
\dot{\psi}^-(z)\end{array}\right)=\left(\begin{array}{ll}
{\cal H}_{11}\ \ {\cal H}_{12}\\
{\cal H}_{21}\ \ {\cal H}_{22}\end{array}\right)\left(\begin{array}{l}
\psi^+(z)\\
\psi^-(z)\end{array}\right)
\ee
where $\psi^\pm(z)=(\bar{z}\vert\psi^\pm>$ and ${\cal H}_{ij}\psi^\pm(z)=
(\bar{z}\vert H_{ij}
\vert\psi^\pm>$. With the generators realization as in eq. (6) and the
factorization of each wavefunction component as in eq. (9) i.e.
\be
\psi^\pm(z)=\prod^{2j}_{m=1}(z-z^\pm_m)\ ,
\ee
we obtain the following equations of motion for the zeros:
\be
\begin{array}{l}
\dot{z}^+_n=i[\omega z^+_n+\frac{\lambda_1\sum\limits^{2j}_{\ell=1}
\prod\limits^{2j}_{\ell\neq m=1}z^{+-}_{nm}+\lambda_2(2jz^+_n
\prod\limits^{2j}_{m=1}
z^{+-}_{nm}-z^{+2}_n\sum\limits^{2j}_{\ell=1}\prod\limits^{2j}_{\ell\neq m=1}
z^{+-}_{nm})}{\prod\limits^{2j}_{n\neq m=1}z^{++}_{nm}}]\end{array}\nonumber\\
{\rm and}\hspace{12.0cm}\\
\begin{array}{l}
\dot{z}^-_n=i[\omega z^-_n+\frac{\lambda_2\sum\limits^{2j}_{\ell=1}
\prod\limits^{2j}_{\ell\neq n=1}z^{-+}_{nm}+\lambda_1(2j\ z^-_n
\prod\limits^{2j}_{m=1}
z^{-+}_{nm}-z^{-2}_{n}\sum\limits^{2j}_{\ell=1}\prod\limits^{2j}_{\ell\neq m=1}
z^{-+}_{nm})}{\prod\limits^{2j}_{n\neq m=1}z^{--}_{nm}}]
\end{array}\nonumber
\ee
where $z^{ab}_{nm}\equiv z^a_n-z^b_m$. Two first comments concerning these
equations of motion are in order: the existence of the spin in the Hamiltonian
coupled to the boson, manifest itself as an interaction between positive $z^+$
and negative $z^-$
zeros; excluding one zero each time the denominators do not cause any small
denominator problem, however in numerical calculations we often had
encountered, in the given machine precision, numerical overflow when some
zeros were approaching very closely.

Assuming $j=1/2$, is the most elementary case where there is only one positive
and one negative zero, which are actually decoupled from each other and they
satisfy separately Riccati equations i.e.
\be
\dot{z}^\pm=i\omega z^\pm -\lambda_{1,2}-\lambda_{2,1}(z^\pm)^2\ .
\ee
In Fig. 1 we plot the motion of the real and imaginary parts of the two zeros
for non-RWA and for RWA.

For the case $j=2$, the interaction of positive and negative zeros appears and
the difference in dynamics between RWA and non-RWA starts to show up. Also
as the next equations show, the complexicity of the system escalades so that
only numerical treatment becomes feasible:
\be
\dot{z}^\pm_1=i\{\omega z^\pm_1+\frac{\lambda_{1,2}(2z^\pm_1-z^\mp_1-z^\mp_2)
+\lambda_{2,1}[2z^\pm_1(z^\pm_1-z^\mp_1)(z^+_1-z^-_2)-z^{\pm 2}_1(2z^\pm_1-
z^\mp_1-z^\mp_2)]}{z^\pm_1-z^\pm_2}\}\nonumber
\ee
\noindent and
\be
\dot{z}^\pm_2=i\{\omega z^\pm_2+\frac{\lambda_{1,2}(2z^\pm_2-z^\mp_1-z^\mp_2)
+\lambda_{2,1}
[2z^\pm_2(z^\pm_2-z^\mp_1)(z^\pm_2-z^\mp_2)-z^{+2}_2(2z^\pm_2-z^\mp_1-z^\mp_2)]}
{z^\pm_2-z^\pm_1}\}
\ee
In Fig.2 we have plotted the dynamics for the real and the imaginary parts of
the positive zeros, in the $(a,b)$, non-RWA and $(e,f)$ RWA; and for the
negative zeros in $(c,d)$ for non-RWA and $(g,h)$ for RWA.

One stricking feature of this dynamics is that to the RWA corresponds an open
trajectory of zeros while for non-RWA dynamics the trajectories are closed.
This might appear odd since one should expect to have more regular paths for
the zeros in the rotating wave approximation, however it can possible be
explained from the fact that the equation of motion
above, are more symmetric in form in the non-RWA case than in the RWA case.

No further results for higher $j$'s will be reported here, since the
complexicity
of the numerical work forbits practically to search the dynamics for some
few tens of $j$ where in accordance to the contraction approximation [23]
the finite-JCM will mimic satisfactory the original model, which in the
non-RWA exibits chaotic behaviour [25]. However we intend to return to this
question elsewhere with some more effective programming routines.
\vskip 1.0 cm
{\bf V. THE JAYNES-CUMMINGS-MODEL CASE}
\vskip 0.5 cm
We finally discuss here the case of the original JCM and derive the
equation of motion for the wavefunction zeros. Based on the Hamiltonian of
the model,
\be
H=\omega a^+a+\omega\sigma_3+\lambda_1(\sigma^+a+\sigma^-a^+)+\lambda_2
(\sigma^+a^++\sigma^-a^-)\ ,
\ee
and using the realizations in eq. (18) [26,27] and the factorization of
eq. (22) we
obtain as in the previous case the following dynamical equation for the
positive and negative zeros:
\renewcommand{\theequation}{\arabic{equation}a}%
\be
\dot{\alpha}^+_n=i\omega\alpha^+_n+i\frac{(\lambda_1+\lambda_2\alpha^+_n)
\prod\limits^\infty_{m=1}E(\frac{\alpha^+_n}{\alpha^-_m})+\lambda_1
\sum\limits^\infty_{\ell=1}\prod\limits^\infty_{\ell\neq k=1}
D^{n;+}_{\ell k;-}}{\sum\limits^\infty
_{\ell=1}\prod\limits^\infty_{\ell\neq k=1}D^{n;+}_{\ell k;+}}
\ee
\addtocounter{equation}{-1}%
\renewcommand{\theequation}{\arabic{equation}b}%
\be
\dot{\alpha}^-_n=i\omega\alpha^-_n+i\frac{(\lambda_2+\lambda_1\alpha^-_n)
\prod\limits^\infty_{m=1}E(\frac{\alpha^-_n}{\alpha^+_m})+\lambda_2\sum\limits
^\infty_{\ell=1}\prod\limits^\infty_{\ell\neq k=1}D^{n;-}_{\ell k;+}}
{\sum\limits^\infty_{\ell=1}\prod\limits^\infty_{\ell\neq k=1}
D^{n;-}_{\ell k;-}}
\ee
where $D^{n;X}_{\ell k;Y}\equiv\left[\frac{\alpha^{X^2}_n}{\alpha^Y_\ell
\alpha^Y_k}-\frac{\alpha^X_n}{\alpha^Y_k}\right]\exp\left(\frac{\alpha^X_n}
{\alpha^Y_\ell}+\frac{\alpha^X_n}{\alpha^Y_k}\right)\ .$

As a mere indication of the time evaluation of the zeros entailed by the
above equations we have plotted in the next figure the case of only one zero.
This is indeed a rather severe oversimplification of the bosonic system,
however even in that case the two existing zeros are interacting with an
exponential non-linearity as the next equation shows,
\renewcommand{\theequation}{\arabic{equation} }%
\be
\dot{\alpha}^\pm=i\omega\alpha^\pm+i(\lambda_{1,2}+\lambda_{2,1}\alpha^\pm)
(1-\frac{\alpha^\pm}{\alpha^\mp})e^{\alpha^\pm/\alpha^\mp}
\ee
In fig. 3 the non-RWA flow of zeros is shown in $(a,b)$ and the RWA in $(c,d)$.
We see first that for the non-RWA case the flow has a nontrivial character
while to the RWA motion, as in the finite case before, corresponds open
trajectories.
Comparing fig. 3 with fig. 2 we find that at least qualitatively the
$SU(2)$-JCM resembles for $j=1$ the RWA evolution of the two zeros in fig. 3.
However this resemblance is rather poor for the case of non-RWA motion.
It will rather require higher values of $j$ to be able to approximate the
non-RWA dynamics of JCM even for only one zero as in fig. 3.
\vskip 1.0 cm
{\bf VI. SUMMARY}
\vskip 0.5 cm
The coherent states as a special basis of vectors in the Hilbert space of
spin and oscillator models map the wavefunctions into analytic functions.
Using standard theorems of complex calculus these analytic functions can be
described in a novel fashion in terms of their roots. This kind of description
has been expanded here to a number of familiar models of quantum optics with
the aim to get an alternative point of viewing the dynamical evolution of these
models. As the most characteristic aspect of this alternative description
we find the appearance of an unexpected high non-linearity hidden in the way
the zeros evolve and interact with each other. This fact impairs to some extend
the
applicability of this method in solving dynamical problems since as we have
seen most of the time only numerical treatment of the ensuing equations is
possible. Despite these difficulties we regard the wavefunctions zeros picture
as a promising one, especially in addressing problems of quantum chaos, as
has been attempted before, since in this picture the quantum mechanical nature
of a given problem is in a way not apparent and one has to deal with a system
of equations of motion of classical mechanics governing the flow of zeros
which however possesses all quantum features of the problem.
\vskip 2.0 cm
\begin{center}
\pagebreak
{\bf ACKNOWLEDGEMENTS}
\end{center}
One of us (D.E.) wishes to thank J. Arponen, R. Gilmore, A.M. Perelomov,
S. Stenholm and K.B. Wolf for useful discussions on zeros.He also likes to
thank
DGICYT (Spain) for financial support.

\pagebreak

{\bf FIGURE CAPTIONS}
\vskip 1.0 cm
Fig.1. For $j=1/2$ the motion of the two zeros is shown for non-RWA in
Fig.1a,b and for RWA in Fig.1c,d.
\vskip 1.0 cm
Fig.2. For $j=1$ the motion of the two positive and two negative zeros is
shown for non-RWA in Fig.2a,c and c,d and for RWA in Fig.2e,f and g,h.
\vskip 1.0 cm
Fig.3. For the truncated JCM the motion of one zero from each component of the
wavefunction (positive, negative) is shown for non-RWA in Fig.3a,b and for
RWA in Fig.3c,d.
\pagebreak


\begin{thebibliography}{99}
\bibitem{1}
J.R. Klauder and B-S. Skagerstam, {\it Coherent States: Applications in
Physics and Mathematical Physics}, (World Scientific, New York, 1985).
\bibitem {2}
E. Schr\"odinger, Naturwissenschaften {\bf 14}, 664 (1926).
\bibitem{3}
R. Glauber, Phys. Rev. {\bf 130}, 2529 (1963); {\bf 131}, 2766 (1963).
\bibitem{4}
J.M. Raddiffe, J. Phys. A {\bf 4}, 313 (1971).
\bibitem{5}
F.T. Arrechi, E. Courtens, R. Gilmore and H. Thomas, Phys. Rev. {\bf A6},
2211 (1972).
\bibitem{6}
A.M. Perelomov, {\it Generalized Coherent States and Their Applications}
(Springer, New York, 1986).
\bibitem{7}
V. Bargmann, Comm. Pure. Appl. Math. {\bf 14}, 187 (1961); {\bf 20}, 1 (1967).
\bibitem{8}
R.P. Boas, {\it Entire Functions} (Academic, New York, 1954).
\bibitem{9}
B.A. Dubrovin and S.P. Novikov, Sov. Phys.- JETP {\bf 52}, 511 (1980).
\bibitem{10}
F.D.M. Haldane and E.H. Rezayi, Phys. Rev. B {\bf 31}, 2529 (1985).
\bibitem{11}
D.D. Arovas, R.N. Bhatt, F.D.M. Haldane, P.B. Littlewood and R. Rammal,
Phys. Rev. Lett. {\bf 60}, 619 (1988).
\bibitem{12}
P. Leboeuf and A. Voros, J. Phys. A {\bf 23}, 1765 (1990).
\bibitem{13}
P. Leboeuf, J. Phys. A {\bf 24}, 4575 (1991).
\bibitem{14}
G.M. D'Ariano, L.R. Evangelista and M. Saraceno, Phys. Rev. A {\bf 45},
3646 (1992).
\bibitem{15}
E. Majorana, Nuovo Cimento [8], {\bf 9}, 43 (1932).
\bibitem{16}
H. Bacry, Comm. Math. Phys. {\bf 72}, 119 (1980).
\bibitem{17}
L.C. Biedenharn and J.D. Louck, {\it Angular Momentum in Quantum Physics},
(Addison-Wesley, Reading, Massachusetts, 1981).
\bibitem{18}
D. Ellinas, Phys. Rev. A {\bf 45}, 1822 (1992).
\bibitem{19}
S. Stenholm and A. Bambini, IEEE Quant. Electon. {\bf 17}, 1365 (1981).
\bibitem{20}
F. Ciocci, G. Dattoli, A. Renieri and A. Torre, Phys. Rep. {\bf 141} \# 1
(1986)
{}.
\bibitem{21}
M.V. Berry, {\it The Diffraction of Light by Ultasound},
(Academic Press, London, 1966).
\bibitem{22}
E.T. Jaynes and F.W. Cummings, Proc. IEEE, {\bf 51}, 89 (1963).
\bibitem{23}
D. Ellinas, Jour. Math. Phys. {\bf 32}, 135 (1991).
\bibitem{24}
J.H. Eberly, N.B. Sanchez-Mondragon and N.B. Narozhny, Phys. Rev. Lett.
{\bf 44}, 1323 (1980);\\
N.B. Narozhny, N.B. Sanchez-Mondragon and J.H. Eberly, Phys. Rev. A {\bf 23},
236 (1981).
\bibitem{25}
P. Milonni, M.L. Shih and J.R. Ackerhalt, {\it Chaos in Laser-Matter
Interactions} (World Scientific, Singapore, 1987).
\bibitem{26}
S.S. Schweber, Ann. Phys. {\bf 41}, 205 (1967).
\bibitem{27}
S. Stenholm, Opt. Comm. {\bf 36}, 75 (1981).
\end{thebibliography}
\end{document}